\documentclass{aa}  

\usepackage{graphicx}
\usepackage{float}
\newcommand{\orcid}[1]{\href{https://orcid.org/#1}{\protect\includegraphics[width=8pt]{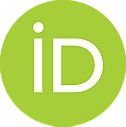}}}
\usepackage{txfonts}
\usepackage{soul}
\usepackage[colorlinks=true,urlcolor=Blue,citecolor=Blue,linkcolor=black,bookmarks=true]{hyperref}
\usepackage[dvipsnames]{xcolor}
\usepackage{ulem}

\newcommand{\vah}[1]{\textcolor{black}{#1}}

\begin{document}

\title{Primordial Binary Stars, Mass segregation and Fractality Effects on the Early Evolution of Young Open Clusters}
\titlerunning{Primordial Binary Stars, Mass segregation and Fractality effects: Early evol. of YOC}
\authorrunning{Amiri et. al.}
   \author{V. Amiri   \inst{1}\fnmsep\thanks{vahid.amiri@uni-heidelberg.de}, 
   F. Flammini Dotti \inst{2,3,1,4}\fnmsep\thanks{ff2415@nyu.edu} \orcid{0000-0002-8881-3078} ,
   X. Pang \inst{5,6}\orcid{0000-0003-3389-2263}
   A.W.H. Kamlah \inst{1,7},
   P. Berczik
\inst{9,10}\orcid{0000-0003-4176-152X}
   B. Shukirgaliyev \inst{11,12,13}\orcid{0000-0002-4601-7065}
   \and
   R. Spurzem \inst{1,8,14}\orcid{0000-0003-2264-7203}
   }

   \institute{Astronomisches Rechen-Institut, Zentrum f\"ur Astronomie der Universität Heidelberg, M\"onchhofstraße 12-14, D-69120 Heidelberg, Germany
\and Department of Physics, New York University Abu Dhabi, PO Box 129188 Abu Dhabi, UAE
\and Center for Astrophysics and Space Science (CASS), New York University Abu Dhabi, PO Box 129188, Abu Dhabi, UAE
    \and
    Dipartimento di Fisica, Sapienza, Universit\'a di Roma, P.le Aldo Moro, 5, 00185 - Rome, Italy
   \and
   Department of Physics, Xi{'}an Jiaotong-Liverpool University, 111 Ren{'}ai Rd., Suzhou Dushu Lake Science and Education Innovation District, Suzhou Industrial Park, Suzhou 215123, P.R. China
   \and
   Shanghai Key Laboratory for Astrophysics, Shanghai Normal University, 
                100 Guilin Road, Shanghai 200234, P. R. China
       \and Max-Planck-Institut f\"ur Astronomie, K\"onigstuhl 17, 69117 Heidelberg, Germany
   \and 
   National Astronomical Observatories, Chinese Academy of Sciences, 20A Datun Rd., Chaoyang District, 100101, Beijing, China
   \and
   Nicolaus Copernicus Astronomical Centre, Polish Academy of Sciences, ul. Bartycka 18, 00-716 Warsaw, Poland
   \and
   Main Astronomical Observatory, National Academy of Sciences of Ukraine, 27 Akademika Zabolotnoho St, 03143 Kyiv, Ukraine
   \and
   Heriot-Watt University Aktobe Campus, 263 Zhubanov Brothers Str, 030000 Aktobe, Kazakhstan 
   \and
   Heriot-Watt International Faculty, K. Zhubanov Aktobe Regional University, 263 Zhubanov Brothers Str, 030000 Aktobe, Kazakhstan
   \and
   Department of Physics, School of Sciences and Humanities, Nazarbayev University, 53 Kabanbay Batyr Ave., 010000 Astana
   \and
   Kavli Institute for Astronomy and Astrophysics, Peking University, 5 Yi He Yuan Road, Haidian District, Beijing 100871, P.R. China
   }

   \date{}

\abstract
{Star clusters form with substructures, which disappear in a relatively short dynamical time scale, and leave behind a smooth density and velocity distributions used in star cluster models. However, star clusters also form with a considerable number of primordial binaries, and primordial mass segregation has been proposed to explain observations of mass segregation in extremely young clusters. 
Current and future observational data may provide better insight into how we observe today formation as a function of cluster age. Numerical simulations can be used to predict and compare primordial mass segregation and the presence of substructures, with the aid of observational data.}
{We want to understand how the combined effect of initial substructure, primordial mass segregation, and primordial binaries affects the dynamical evolution of the cluster, and which one of these features is the most important to agree with observations. 
}
{We use Nbody6++GPU to simulate the dynamics of star clusters with initial substructure, primordial mass segregation, and primordial binaries, and we also study the relative importance of the processes. Initial models were generated by a modified version of  \textsc{McLuster}, and we compared our results with observational data from \citealt{pang22} database of open clusters.}
{Our results show that primordial mass segregation and binaries do not change the result already obtained in previous works, as the time scale on which initial substructure disappears is of the order of few Myrs. However, we also find that in the presence of initial substructure, primordial mass segregation does not lead to an early expansion of the cluster. The processes in the core, discussed in previous works, lead to a loss of low mass stars and early expansion, are postponed in the presence of initial substructure. Finally, we find from comparison with observed clusters that primordial mass segregation is not a fundamental process to reproduce observational data.}
{}


   \maketitle
%

\section{Introduction}
\label{intro}
Stars are believed to form from molecular clouds in regions containing high-density gas (\vah{\citealt{lada2003, zinnecker2007, kruijssen,lomax}} and \citealt{chev}). These clouds tend to contract as the gravitational force within them overcomes the internal pressure once the cooling process begins. As a result, the gas collapses locally, and stars form. The contraction process can create filaments that form a network that feeds new materials into the nodes (\vah{\citealt{andre2010, krause2020, brooke2025}} and \citealt{laverde2025}). Star clusters form smaller substructures that contain gas and grow by absorbing the gas from the giant molecular cloud surroundings (\vah{\citealt{fujii2022}} and \citealt{karam}). In the traditional understanding of cluster formation, a three-step process is described: (i) a cold local substructure, which is named cloud, in the giant molecular cloud environment, begins to contract; (ii) this triggers the formation of protostellar cores along collapsing stellar filament; and (iii) massive formed stars emit radiation that ablates the remaining gas, leaving behind a relatively dense cluster of stars, known as young massive clusters (\citealt{kroupa01} and \citealt{zwart10}).

\par Observational data argue that young (with age  $\leq 100$ Myrs) open clusters are often associated filamentary substructures (\vah{\citealt{krause2020,beccari2020, pang22,greg2024}} and \citealt{arnold2024}), which are the remnants of filaments. However, the initial signs of substructures in morphology and kinematics disappear after several crossing times ( see \citealt{binney} for the definition), and observations of clusters showing sub-group merging support this theory (\citealt{kuhn}).

\par Some Young star clusters have their most massive stars concentrated in the center of the star cluster, which can be considered a sign of primordial mass segregation (\citealt{deg, gou} and \citealt{bonnell2006}). The origin of mass segregation remains unclear, as it could be either primordial (i.e., caused by the local formation of stars in the giant molecular cloud) or dynamical (i.e., caused by the dynamical evolution of the star cluster). In the former case, the star cluster forms with the most massive stars concentrated at the center, while in the latter case, the most massive stars are shifted toward the center by two-body interactions after formation. Determining the origin of mass segregation could help distinguish between models of massive star formation, in particular, whether the mass of the core in which they form or a favorable position in the star cluster determines the masses of the most massive stars (\citealt{bonnell2006} and \citealt{krum}). If mass segregation is primordial, it could support competitive accretion since massive stars should form in the center of the star cluster. Previous studies have shown that dynamical mass segregation does not justify the observations in young clusters, such as the Orion Nebula cluster (\citealt{bonn98}). 

 \par Although many studies ignore the primordial binary systems in simulations, observational data show a large variation for the fraction of binary systems in star clusters, ranging from 0.2 to 0.5 (cf. e.g. \citealt{milone2012, li2013} and \citealt{ramirez2024}), or in some cases even higher ( e.g. \citealt{abt,duq,leinert93} and \citealt{offner23}). Some studies (e.g. \citealt{marks12}) reported that this fraction is close to 1.0 for massive stars in star clusters.

\par Recent numerical studies have examined the impact of primordial mass segregation,  degrees of primordial fractality, and fractions of primordial binary stars. Some studies suggest that mass segregation may be primordial in star clusters (\citealt{haghi} and \citealt{alfaro}). Primordial mass segregation can account for the low mass star depletion in the mass function  (\citealt{bamg}) and the early expansion of the clusters (\citealt{vesperini09}).

\par  Numerical simulations also show that star clusters are formed by the merging of sub-stellar groups (\citealt{goodwin, mcmillan2007, alis, laverde2025} and \citealt{brooke2025}). Specific research indicates that primordial fractality may be present during the early stages of cluster formation (\citealt{ballone} and \citealt{laverde2025}), and affects the overall expansion of star clusters (\citealt{tor}). However, initial substructures tend to be smooth and centrally concentrated at the very early stage of the evolution of the cluster (\citealt{parker14, sills2018, daff, greg2024} and \citealt{brooke2025}).

\par The presence of substructure in the star cluster is one of the most significant factors in the dynamical evolution of binary stars (\citealt{kupper, parker11b, parker14, dorval17, daff} and \citealt{tor}). In addition, Various degrees of fractality and the fraction of binary stars affect the degree of mass segregation in clusters (\citealt{kupper, parker11, parker14, parker2} and \citealt{parker22}), even though some results indicate that clusters with primordial mass segregation and primordial binary stars are more compatible with some \vah{observed clusters} (e.g. \citealt{yu} and \citealt{pav}).
   
\par From the last decades, researchers have been highlighting the effect of primordial binary systems on the evolution of star clusters (e.g. \citealt{kroupa95, kroupa99,reip, kacz, wangdragon, dragonarca2, dragonarca3, cour} and \citealt{cour2024}) as they are the energy source in the clusters, preventing infinite collapse and resulting in high impact on the long-term evolution of different features of the clusters. Some results suggest that the dynamical evolution of primordial binary stars affects mass inflation of the star cluster (\citealt{rastello}), and leads to the reaching of dynamical equivalencies (\citealt{belloni2}).  \citealt{wang22} argued that, in the presence of a black hole,  the dynamical evolution of core and half mass radii ($r_h$) only depends on massive binary stars.

\par Although many studies have explored how different properties affect the evolution of star clusters, a comprehensive analysis that includes primordial mass segregation, primordial fractality, and the initial fraction of binaries is still needed. Such an analysis allows us to compare different models in detail and evaluate not only the effect of each feature on its own but also the combined influence of these features on cluster evolution. In this work, we studied the dynamical evolution of young open clusters while considering two extreme levels of primordial mass segregation and primordial fractality, both with and without primordial binary systems. This approach provides a more complete picture of their evolution under various starting configurations. We also investigated whether primordial mass segregation could arise from primordial fractality in the clusters, and calculated the timescale over which substructures disappear when primordial binaries and primordial mass segregation are present. 

The paper is structured as follows: Section 2 covers the methodology and initial conditions. The simulation results for clusters with various features are presented in Section 3. Section 4 then discusses the conclusion and summary.
\section{Methodology and Initial Conditions}

We have simulated eight star cluster models using the state-of-the-art direct force integration code \textsc{NBody6++GPU}, which is optimized for parallelization using simultaneously three different levels, at the bottom-end many-core GPU-accelerated parallel computing (\citealt{Nitadori2012} and \citealt{wanggpu}), mid-level OpenMP thread-based parallelization, and upper-level MPI parallelization (\citealt{spurzem99}). It yields excellent sustained performance on current hybrid massively parallel supercomputers (\citealt{Spurzem2023}). It is a successor to the many direct force integration $N$-body codes of gravitational $N$-body problems, which were originally written by Sverre Aarseth (\citealt{Aarseth1999, Aarseth2003}).  The code can numerically evolve up to few million bodies (see \citealt{wangdragon, dragonarca2, dragonarca3}).

\begin{figure*}[h]
\centering
\hspace{-0.6cm}
\includegraphics[height=20.5cm,width=16cm]{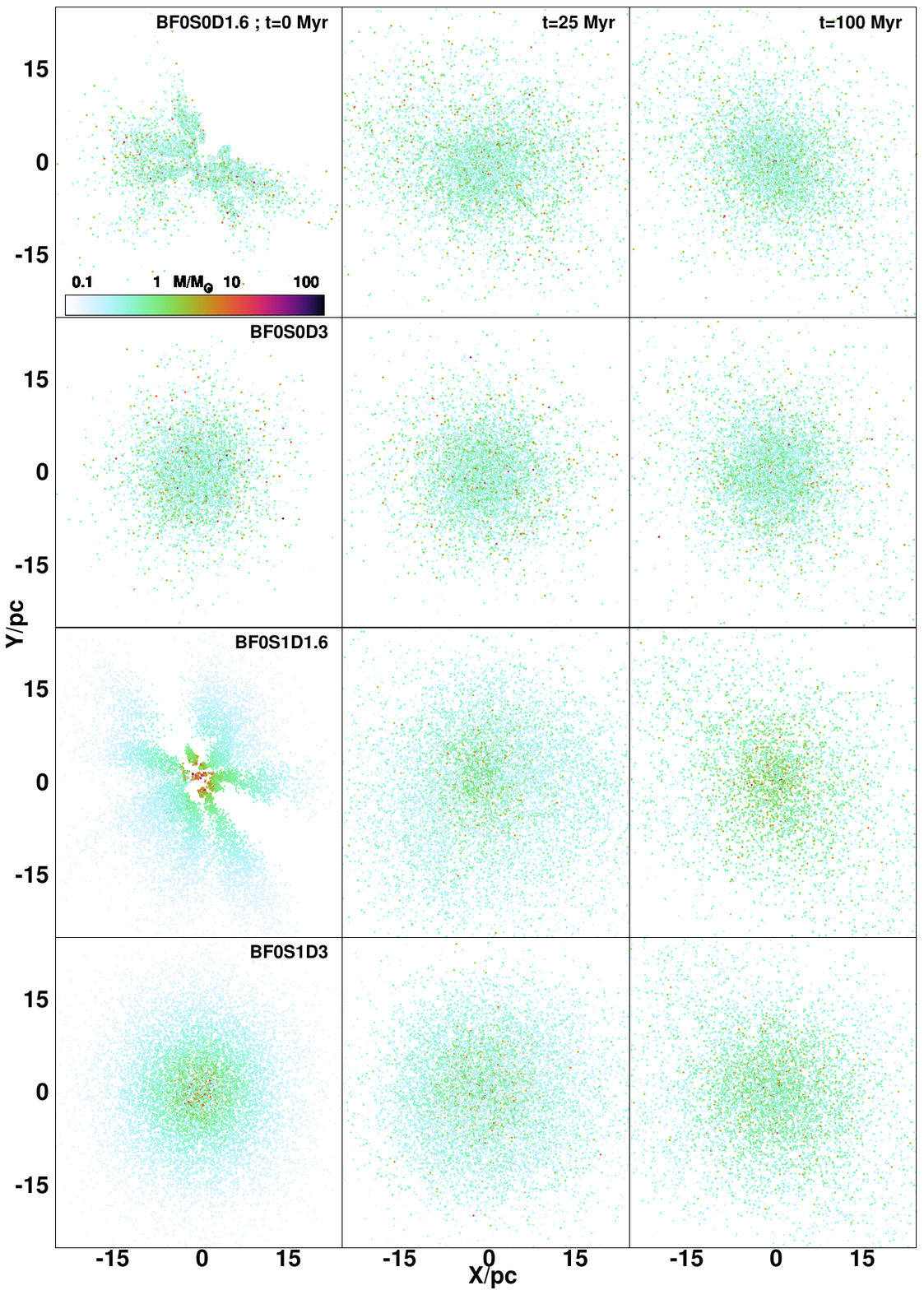}
 
    \caption{Face-on initial stellar distribution for \vah{models with no binary systems in three different time steps in the evolution}. Different colors indicate varying masses, as shown by the color bar.}
    \label{ini}
\end{figure*}

\par We set up our models with similar initial conditions, except for the properties listed in Table \ref{table}. Implementation of the initial condition is done using the most updated version of the code \textsc{McLuster} (\citealt{kroupa08,kupper} and \citealt{agostino}). This code is free and open-source\footnote[1]{https://github.com/agostinolev/mcluster} .

\begin{table}[h]
\centering
\scalebox{0.75}
{
    \begin{tabular}{|c||c|c|c|c|c|c|c|}

        \hline 
        \textbf{Name} & \textbf{BF} & \textbf{S} & \textbf{D} & \textbf{M}$_{tot}$ \tiny $[\times  10^3 M_\odot]$ & \textbf{T}$_{rh}$ \tiny $[Myr]$ & \textbf{T}$_{ch}$ \tiny $[Myr]$ & \textbf{N} \tiny $[\times  10^3]$ \\
    
        \hline
        \tiny \textbf{BF0S0D1.6} & $0.0$ & $0$ & $1.6$ & $5.83 $  & $607.41$ & $4.47$ & $10$ \\
        \hline
       \tiny \textbf{BF0S0D3} & $0.0$ & $0$ & $3.0$ & $5.83 $  & $671.37$ & $8.95$ & $10$\\
        \hline
          \tiny \textbf{BF0S1D1.6} & $0.0$ & $1$ & $1.6$  & $5.83 $ & $607.43$ & $4.47$ & $10$ \\
        \hline
          \tiny \textbf{BF0S1D3} & $0.0$ & $1$ & $3.0$ & $5.83 $ & $671.37$ & $9.27$ & $10$\\
        \hline
          \tiny \textbf{BF0.1S0D1.6} & $0.1$ & $0$ & $1.6$  & $6.41 $ & $670.78$ & $6.40$ & $11$\\
        \hline
          \tiny \textbf{BF0.1S0D3} & $0.1$ & $0$ & $3.0$ & $6.41 $ & $701.27$ & $8.84$ & $11$\\
        \hline
          \tiny \textbf{BF0.1S1D1.6} & $0.1$ & $1$ & $1.6$  & $6.41$ & $670.78$ & $6.10$ & $11$\\
        \hline
          \tiny \textbf{BF0.1S1D3} & $0.1$ & $1$ & $3.0$  & $6.41 $ & $640.29$ & $8.84$ & $11$\\
        \hline
    \end{tabular}
}
    \caption{Initial conditions for different models. The columns give the names of the models, the primordial fraction of binary systems (BF), the primordial mass segregation factor (S), the degree of primordial fractality (D), the total mass, the half-mass relaxation time, the half-mass crossing time (see \citealt{binney} for the definitions) of the star cluster, and the total number of particles in the models.}
    \label{table}
\end{table}

\par \vah{The simulations are initialized at a stage where the clusters have already reached virial equilibrium, with no gas}. The \vah{metallicity of stars is 0.0005 (following \citealt{kamlah})}, and the density distribution is the King model (\citealt{king}), using $W_0=6$. \vah{The initial half-mass radius of $ r_h = 5pc $ is adopted to ensure that dynamical and stellar evolutionary processes occur on comparable timescales, allowing their coupled interplay to be properly captured.} These star cluster models are in virial equilibrium, and the initial mass function (IMF) follows \citealt{kroupa2001}, with stellar mass ranging from $0.08$ $M_\odot$ to $150.0$ $M_\odot$. The star clusters are subjected to the standard solar neighborhood tidal field.

 \par Determining the binary fraction in star clusters is still a challenge. As mentioned in Section \ref{intro}, observations show a large variation from cluster to cluster, ranging from a fraction of 0.2 up to 1.0. Generally, the number of binaries drops with time, since soft binaries are subject to disruption by encounters with other stars or other binary systems.  Soft binaries also may be formed in clusters via three-body interactions (\citealt{cour2024}). \vah{Observational studies based on Gaia DR3 data by \citealt{cordoni2023} indicate that the primordial binary fraction in open clusters ranges from 0.15 to 0.6. \citealt{Jiang2024} reported binary fractions in the range 0.06-0.34 (for systems with $q = \frac{m_{2}}{m_{1}} \geq 0.5$, where $m_{1}$ and $m_{2}$ are the masses of the primary and secondary stars, respectively) with a median value of 0.17. Furthermore, \citealt{pang2026} showed that most primordial soft binaries, which dominate the binary population of open clusters, dissolve within $\sim$ 5-8 Myr. Motivated by these observational and theoretical results, we adopt the primordial hard binary fraction of 0.1 in our simulations to compare cluster evolution with models that do not include primordial binaries. In this study, the binary fraction refers to the fraction of stars in the cluster that belong to binary systems. Binary pairing is performed following the prescription of \citealt{sana}.}
  
 \par \vah{Regarding the period distribution, we used the \citealt{kroupa95} period distribution for M $<$ 5M$\odot$; and \citealt{sana}, and \citealt{oh15} period distribution for M $>$ 5M$\odot$. Using the option implemented in the \textsc{McLuster} code, we restricted the maximum semi-major axis such that only primordial hard binaries were included in the initial models.}

 \par Initial fractal substructures were implemented using the method described in \citealt{goodwin}, \citealt{kupper}, and \citealt{agostino}. The degree of fractality is determined by the fractal dimension D. In this study, we investigated two extreme values: D$= 3.0$, which results in no substructure, and D$= 1.6$, which results in a cluster with the maximum degree of fractality. 
 
 \par Finally, primordial mass segregation is implemented following the methodology described in \citealt{bamg} and \citealt{kupper}. The degree of segregation is controlled by a parameter S, where S$= 0$ results in a random distribution (and thus, the standard density distribution profile), and S$= 1$ produces full primordial mass segregation; two extreme cases that we will explore in the following sections.
\section{Results}
\par Here, we analyze the results of the simulation of the evolution of our models using the different characteristics listed in Table \ref{table}. Figure \ref{ini} shows the initial distribution of the stars in the clusters with no primordial binary systems on the face-on plane. \vah{Figure \ref{ini} presents the initial distribution of stars in clusters without primordial binaries, shown in the face-on projection, together with snapshots at t = 25 Myr and at the final stage of the simulations. The color bar indicates the corresponding stellar mass range. For models with primordial fractal structure, the initial distribution is consistent with the setups of \citealt{kupper} and \citealt{rossi}. In the following sections, we will discuss the dynamical evolution of all models over the course of the simulations.}

\par We compared our simulation results with 65 observed young open clusters with ages  $\le 100$ Myrs from \citealt{pang22} and \citealt{pang23}. In this sample, the maximum distance of clusters from the observer is $=652$ pc. The young open cluster sample only contains stars that are brighter than magnitude $=21$ mag, the Gaia observational limit (\citealt{gaia}), with an average completeness mass limit of $=0.3 M_\odot $ (\citealt{pang24}). When it comes to comparison with observational data, we applied the same magnitude and mass limit to our simulated models. Note that we replaced the center-of-mass features of binary stars when comparing the results to the observations. The age of young open clusters is derived via isochrone fitting (\citealt{pang22}). Since there is a fraction of clusters younger than $30$ Myrs, the uncertainty of isochrone fitting for young age goes up to $20 \% $ maximum (\citealt{bressan2012}), while it is smaller for older clusters. In this work, we considered the 20\% age uncertainty for the age of observed young clusters as an upper limit. \vah{In addition, we compute sliding-window averages of the observational data for the different parameters considered in this study, allowing us to trace their overall trends and general behaviour.}

\subsection{General Dynamical Evolution}
 \label{gen}

\begin{figure*}[h]
\hspace{-0.6cm}
\includegraphics[scale=1.5]{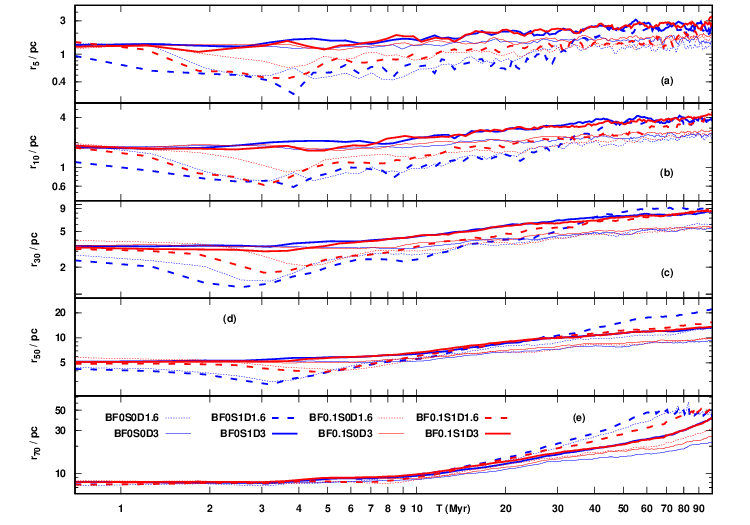}
 
    \caption{Evolution of different Lagrangian radii, 5\% (panel a), 10\% (panel b), 30\% (panel c), 50\% (panel d), and 70\% (panel e), for all models. Red lines depict models with primordial binary stars, while blue lines depict clusters without primordial binary stars; dashed lines represent models with primordial fractality, and solid lines represent clusters without substructures; and thick lines are used for models with primordial mass segregation, while thin lines are used for those without primordial mass segregation.}
    \label{lagr}
\end{figure*}

\begin{figure*}[h]
\hspace{-0.6cm}
\includegraphics[scale=1.5]{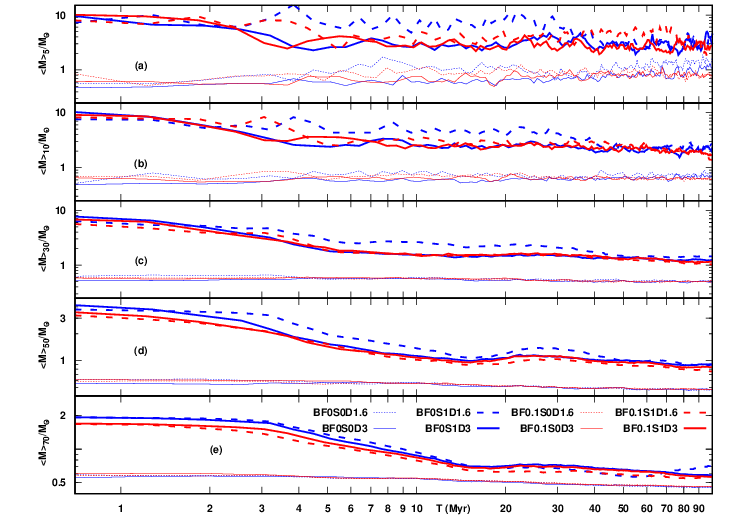}
 
    \caption{Averaged mass within the different Lagrangian radii,  5\% (panel a), 10\% (panel b), 30\% (panel c), 50\% (panel d), and 70\% (panel e), for all models. The specifications of the lines are the same as in the Figure \ref{lagr}.}
    \label{lagr-m}
\end{figure*}

\par The dynamical evolution of Lagrangian radii is often used to study the global evolution of the star cluster in the inner and outer regions. The averaged mass in different shells reveals, instead, the evolution of mass segregation in the cluster. The Lagrangian radius is defined as a star cluster shell that contains a certain percentage of the total mass of the star cluster. In this study, the mass inside the Lagrangian radius is calculated based on the initial total mass of the clusters. Figure \ref{lagr} presents the evolution of the Lagrangian radii for all eight cluster models, and Figure \ref{lagr-m} shows the evolution of the average mass within the various Lagrangian radii.
 
\par Figure \ref{lagr} illustrates that models without primordial fractality show a gradual and consistent increase in their Lagrangian radii throughout the evolutionary period, as compared to non-fractal models. The behavior of the Lagrangian radii, in the inner regions, exhibits two phases in models with a high degree of fractality; the cluster undergoes a contraction before 5 Myr and then expands. The models show high-density sub-regions where, on relatively short time scales, some lighter stars gain energy from the two-body interactions within the considered sub-region and move into the low-density regions. Consequently, after merging these substructures, the cluster gains more mass in the inner regions compared to other models after the disappearance time of the primordial substructures (see section \ref{fract}), and it leads to a decrease in the Lagrangian radii in the inner parts of the cluster. The presence of primordial binary stars mitigates the contraction of the inner part of the clusters, as they act as energy sources in the star clusters (\citealt{binney}).

\par In models with primordial mass segregation, low-mass stars gain energy through interactions with massive stars and move toward the outer regions, expanding the outer regions of the star clusters. In addition, the binding energy is larger in the cluster center, where mass loss from stellar evolution is more dominant (due to heavy masses already standing in the inner regions). This will eventually result in a rapid expansion of the clusters (\citealt{vesperini09} and \citealt{krause2020}). Nevertheless, this expansion is postponed when the cluster has an initial degree of fractality. Even in the presence of primordial mass segregation, inner regions did not exhibit expansion before roughly $30$ Myrs and contracted at the early stages of the simulations.

\par Models with primordial mass segregation show higher average mass in Lagrangian radii than models that were initially not segregated, as shown in Figure \ref{lagr-m}. However, the interesting result is that, at the \vah{time period of $\sim 3-10$ Myr in the simulations}, models with primordial fractality exhibit slightly higher values for the averaged mass in their inner regions compared to other models. The idea that the dynamical evolution and merging of primordial substructures strengthen mass segregation, particularly in the early phases of star cluster evolution (\citealt{mcmillan2007} and \citealt{alis}), is supported by the larger average mass \vah{in the inner regions ($r_{5},\ r_{10}$, and $r_{30}$)} for models with primordial fractality as compared to models without primordial substructures.

 \subsection{Mass Loss and Fraction of Escapers}
 \begin{figure*}[h]
\hspace{-0.6cm}
\includegraphics[scale=1.5]{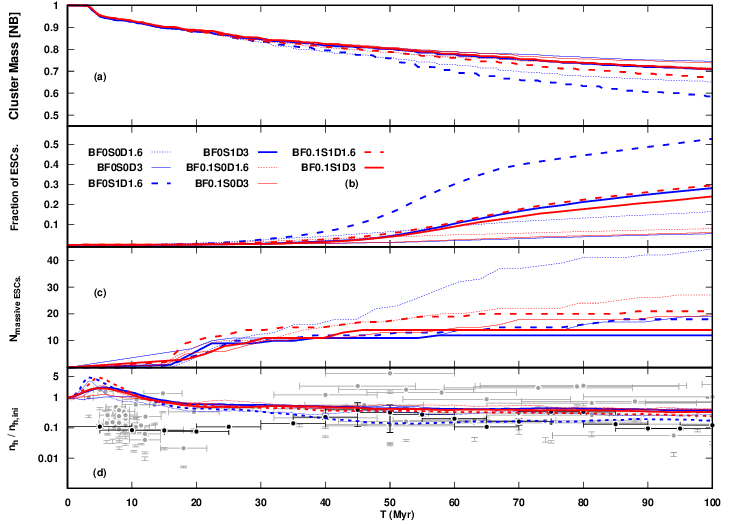}
 
    \caption{The total mass (panel a), the fraction of escapers (panel b), the number of massive ($M > 3 M_{\odot}$) escapers, (panel c), and the ratio of the number density in the $r_h$ (half-light radius for the observational data) to its initial value (\vah{the mean density of the five youngest clusters} in the observational data sample) (panel d), for all models. The observational data (\citealt{pang22}) are shown in grey, while the black points indicate the values averaged using the sliding-window method.}
    \label{mass}
\end{figure*}

\par In this section, we focus on how the combinations of different characteristics impact mass loss in our models, as shown in Figure \ref{mass}. Panels (a), (b), (c), and (d) of the Figure show the total mass, the fraction of escapers \vah{(particles that reach a distance more than twice the tidal radius of the cluster)}, the number of massive ($M> 3 M_{_\odot}$) escapers, and the changes in number density in $r_h$ for all models. 

\par According to the panels (a) and (b) of the Figure \ref{mass}, a larger mass loss of the star cluster is expected in models with primordial fractality and mass segregation. Primordial mass segregation increases the likelihood of close interactions between massive stars and light stars in the inner regions of the cluster. This larger probability of encounters allows lighter stars to gain energy, causing them to be ejected from the star cluster. Furthermore, in the inner regions of the clusters, primordial mass segregation leads to further mass loss due to stellar evolution (\citealt{vesperini09} and \citealt{krause2020}). 

\par In the models with fractality, local dense regions within the star clusters increase the likelihood of two-body interactions, raising the probability of lighter stars acquiring additional energy compared to other models. Consequently, these stars migrate from the local denser regions to regions with shallower gravitational potential, thus facilitating their escape from the star cluster. Furthermore, clusters including substructures result in a larger binary system dissolution rate, which increases the number of escapers (\citealt{dorval17}; see also the section \ref{binary}).  The number of escaping stars is influenced by the presence of primordial binary stars, as shown in Figure \ref{mass}. Primordial binary stars mitigate the increase in the number of escapers from the star clusters by cooling the core of the cluster.

 \par Sub-regions create areas of high density with a short relaxation time scale, leading to more frequent encounters. In this scenario, massive stars can also gain energy and migrate to more dispersed regions. Once the substructures dissipate, the massive stars are more likely to escape the cluster. This results in an increased number of escapers among massive stars (\vah{\citealt{fujii2011science,perets2012,gavagnin}} and \citealt{krause2020}), particularly when they are part of wide dynamical binary systems. However, binaries within these sub-regions can release energy into their systems, reducing the likelihood of encounters. As shown in panel (c) of Figure \ref{mass}, our study indicates that while the model \textit{BF0S0D1.6} has fewer escapers compared to all other models with primordial mass segregation, it experiences a higher mass loss due to the presence of more massive escapers. 
 
\par Panel (d) of Figure \ref{mass} shows the ratio of number density in $r_h$ (which is the half-light radius for plotted observational data) to its initial value (\vah{the mean density of the five youngest clusters} in the observational data sample). \vah{In this panel, the observational data are shown in grey, while the black points represent values averaged using a sliding-window method, which highlights the overall trend of the data and suppresses local fluctuations.} At young ages, the observed value of the number density ratio is lower than in our simulations, which could be associated with incomplete membership of stars from observations due to a magnitude limit or different star formation environments producing clusters with different stellar densities. However, discussion of the latter is out of the scope of the current paper. After $\approx 10$ Myr, however, our models show a decline in density that brings them into good agreement with the observed clusters. In particular, for clusters older than $\approx 20$ Myr, our simulations follow the same evolutionary track as the observational data.
 \subsection{Evolution of Binary Fraction}
 \label{binary}
 \begin{figure*}[h]
\hspace{-0.6cm}
\includegraphics[scale=1.5]{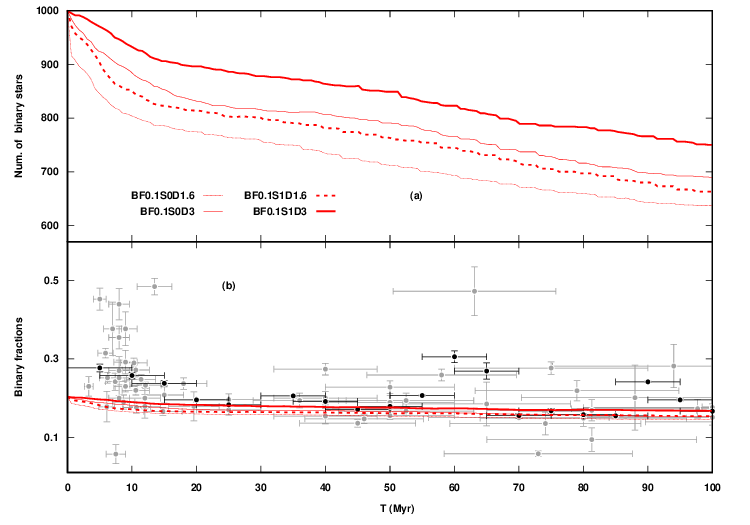}
 
    \caption{Evolution of the total number of binary systems (panel a) and the comparison of fractions of binary stars with observational data (\citealt{pang22}) (panel b) for models that include primordial binary systems. \vah{The observational data are shown in grey, while the black points indicate the values averaged using the sliding-window method.}}
    \label{nb}
\end{figure*}

\par Within the cluster, binary stars act as an energy source and postpone or stop core collapse (\citealt{binney}) and strongly impact the dynamical evolution of star clusters. Panel (a) of the  Figure \ref{nb} shows the evolution of the number of binary stars, and panel (b) compares the observable binary fraction in our models with observational data over the time of simulations. In this study, we restrict our models to an initial hard binary fraction of $10\%$. Wide (or soft) binaries, which dominate the observed binary population at early times, are not included in our initial conditions but will be addressed in future work, as well as the higher fraction of initial binary systems.

\par Panel (a) shows that in the first few Myrs of evolution, the model with primordial fractality reveals a significant decline in the number of binary stars. Star and binary star interactions occur more frequently in high-density sub-regions (\citealt{dorval17}). Thus, several of the binary stars dissolve during the time that the substructures disappear. 

\par When primordial mass segregation is present in the cluster, interactions between massive stars and the binary stars are less common in the outer regions of the cluster, which results in lower binary dissolution; however, when primordial mass segregation is absent, these interactions can occur throughout the cluster, leading to higher binary system dissolution. 

\par Panel (b) of the Figure \ref{nb} compares the fraction of binary stars of the simulated and observed star clusters. \vah{Due to selecting only observable binary systems, the primordial binary fraction increases to 20\%. The observed binary fraction in young clusters shows a significant decrease from $\sim$ 50\% to $\sim$ 20\%, which has been interpreted as evidence for early disruption of soft binaries (\citealt{pang23} and \citealt{pang2026}). However, when the data are averaged using a sliding-window method (black points), this apparent decline is no longer evident}. For our models, which include only hard primordial binaries, only a small decrease is found. After $10$ Myrs, the observable binary fraction of our models generally reaches an agreement with observations.

 \subsection{Mass Segregation Evolution}
\begin{figure*}[h]
\hspace{-0.6cm}
\includegraphics[scale=1.5]{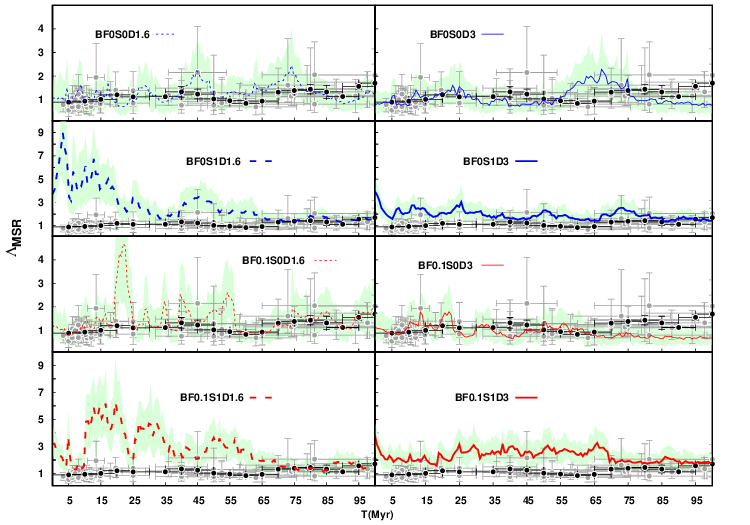}

\caption{Evolution of $\Lambda _{MSR}$, for all models compared to this value for observed clusters (\citealt{pang22}). The parameter $\Lambda _{MSR}$ is calculated using various random sets of stars, and all possible values are displayed in the green area. \vah{The observational data are shown in grey, while the black points indicate the values averaged using the sliding-window method.}}
\label{seg}
\end{figure*}

\par To identify the degree of mass segregation in our models, we followed \citealt{mst} method, which compares the {\it minimum spanning tree}, MST\footnote[1]{An MST connects all the stars (as nodes in a graph) without loops between nodes, minimizing the total length of edges.} of the top N most massive stars (NMST \footnote[2]{Given that observational data from clusters contain only tens to hundreds of stars, and as choosing the NMST value is arbitrary (\citealt{mst}), we used NMST $=10$ for our study.}) with those of a random sample of stars to produce a quantitative measure of mass segregation. \vah{The degree of mass segregation, $\Lambda_{MSR}$, is determined by comparing the average MST length of randomly selected stars with the MST length of the most massive stars.} The value of $\Lambda _{MSR}$ corresponds to the different mass segregation phases. $\Lambda _{MSR} \simeq 1$ shows the random distribution of stars (i.e., no mass segregation), $\Lambda _{MSR} > 1$ shows there is a degree of mass segregation, and  $\Lambda _{MSR} < 1$ shows inverse mass segregation (massive stars are in outer regions of the cluster and lighter stars occupy the inner areas). Clearly, models with primordial mass segregation will have higher values for $\Lambda _{MSR}$ as compared to non initially segregated models.  We note that different NMST values result in different values for $\Lambda _{MSR}$ (\citealt{Allison10, parker2, Hetem} and \citealt{parker22}). The mass range of the stars in the sample is relatively broad when the value of NMST is large. Consequently, significant numbers of the massive stars in the sample have masses nearly identical to those of the stars in the random sample, resulting in values of $\Lambda _{MSR}$ that are quite near to one. The larger the NMST, the closer $\Lambda _{MSR}$ is to one. 

\par The evolution of the mass segregation parameter $\Lambda _{MSR}$ for all models and its comparison with the observational data are shown in Figure \ref{seg} \vah{(the observational data are displayed in grey while the values averaged using the sliding-window method are shown by the black points)}. When accounting for the various samples of observable stars, the green-shaded areas display every possible value for $\Lambda _{MSR}$ (same for the observational data point vertical error bars) while the mean values are shown as lines. 

\par The value of $\Lambda _{MSR}$ is influenced by the dynamical phenomena of the cluster over short time scales, but the overall trends provide a useful comparison. Models without primordial segregation exhibit a better agreement with the observational data, as seen in Figure \ref{seg}. Primordially segregated models, on the other hand, result in larger values of $\Lambda _{MSR}$ compared to observed clusters, especially in the early simulation phases. Over time, however, $\Lambda _{MSR}$ in these models gradually decreases as the massive stars evolve into compact objects.

\par For models with primordial fractality, in substructures, due to shorter time scales, massive stars sink to the center of these sub-regions. This dynamical segregation can be inherited in the final structure of the cluster after the disappearance time of the substructures (\citealt{mcmillan2007}). In our models, in general, models with primordial fractality have slightly higher values for $ \Lambda _{MSR} $ than models without primordial fractality, as the figure illustrates. This supports the idea that mass segregation is accelerated in the presence of substructures and can occur dynamically at the first stages of the evolution of star clusters (\citealt{mcmillan2007} and \citealt{brooke2025}). However, when massive stars are in the central region, due to the high probability of encounter, they can be ejected from the area or even the cluster (\vah{\citealt{brooke2025}}). Due to this, the mass segregation lasts for a short time, as can be seen in the figure. 

\par \vah{In our models, primordial binary systems generally do not have a significant impact on the overall evolution of mass segregation. However, in models with primordial fractality, they produce strong but short-lived mass segregation signals between approximately 10 and 60 Myr of evolution. As mentioned earlier, the merging of substructures drives massive stars and, in models with primordial binary systems, also drives massive binary systems toward the cluster core. Once in the core, interactions between massive binaries and lower-mass stars transfer energy to the lighter stars, pushing them to larger radii and further enhancing the segregation signals. However, these interactions also increase the probability of ejecting massive stars from the core. This disrupts the segregated structure and rapidly erases these signals.}

 \subsection{Evolution of Substructures}
 \label{fract}

\begin{figure*}[h]
\hspace{-0.6cm}
\includegraphics[scale=1.5]{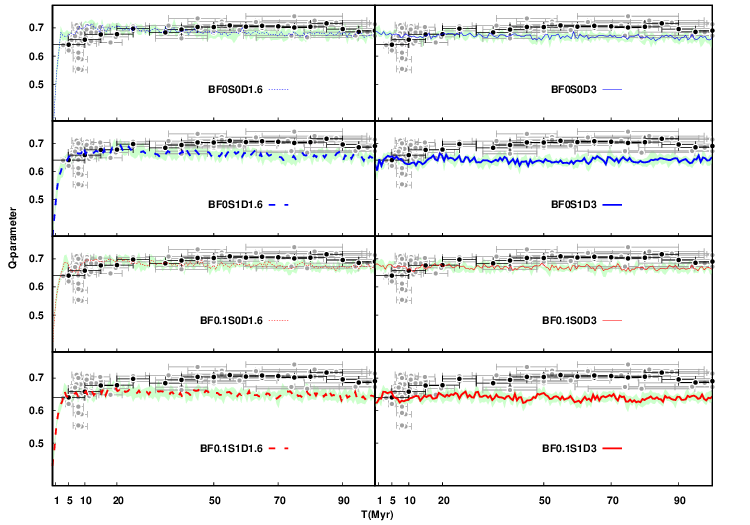}

\caption{Evolution of Q-parameter for all models compared to this value for observed clusters (\citealt{pang22}). When various random sets of stars are applied to calculate the Q-parameter, all possible parameter values are displayed in the green area.  \vah{The observational data are shown in grey, while the black points indicate the values averaged using the sliding-window method.}}
\label{frB}
\end{figure*}

\par The evolution of substructures is important for the long-term evolution of the clusters. \citealt{parker14, sills2018, greg2024} and \citealt{laverde2025} stated that the disappearance time scale of the substructures is very short. In this work, we examined the impact of primordial binary systems in addition to the effect of primordial mass segregation on the evolution of substructures in our models to see how this timescale changes. 

\par To study the evolution of the fractal substructures, we employed the MST method to calculate the Q-parameter in 3-D, which allows us to determine the degree of fractality during the dynamical evolution of the star clusters (\citealt{cart} and \citealt{ballone}). We calculated the Q-parameter in a sphere with a radius of $r_{h}$ for our models and the half-light radius for the observational data. \vah{We employ the MST method to calculate the Q parameter to determine the degree of fractality (\citealt{cart} and \citealt{ballone}). The Q parameter is Q $= l_{MST} / l_{cg}$, where $l_{cg}$ represents the mean length of the edges of the comprehensive $3$-D graph of the stars in the region, formed by connecting each star to every other star (as the nodes of a graph), and $l_{MST}$ is the mean length of the edges of the MST of all stars in the region. Due to their disparate natures, these two measures are inherently scaled differently. Both metrics must be normalized concerning the size of the region. Firstly, $l_{cg}$ is normalized by the region's radius. Second, $l_{MST}$ is normalized to the volume of the region  (\citealt{cart} and \citealt{ballone}).} Clusters with Q-parameter $> 0.7$  exhibit a radial distribution, while clusters with Q-parameter $< 0.7$ exhibit a higher degree of fractality. Q-parameter $ \simeq 0.7 $ shows that the cluster does not show signs of fractality (\citealt{ballone}). 

\par Figure \ref{frB} reveals the evolution of the Q-parameter for all models during the time of simulations \vah{(The grey and black dots in this figure represent the observational data and the values averaged using the sliding-window approach, respectively.)} For all models, we calculated the Q-parameter using sets of randomly chosen stars in the region, and green areas display all possible values for the Q-parameter, while the lines display the mean value. As the Figure \ref{frB} shows, all models with primordial fractality are in better agreement with observational data compared to models without any initial degree of fractality during the first few Myr of evolution. \vah{This better agreement arises because observations indicate that many very young star clusters (with ages $<10$ Myr) exhibit a significant degree of substructure, characterized by low values of the Q-parameter. In our simulations, although the initial fractal structure dissipates rapidly, the models that include primordial fractality reproduce a comparable level of substructure during the first $\sim5$ Myr, consistent with the observational constraints. This early structural similarity is important, as the initial degree of fractality can influence the subsequent dynamical evolution of the clusters.}

\par We showed that primordial fractality disappears after $\approx 5$ Myrs and becomes similar to a smoothly distributed system. This is in agreement with the results reported by \citealt{alis}, \citealt{parker14, sills2018, parker22, greg2024} and \citealt{laverde2025}. This disappearance time range is almost the same as the time of the contractions in the inner regions of the models with primordial fractality (see Figure \ref{lagr} and \ref{lagr-m}). According to our results, the time scale of the disappearance of the substructures is not significantly impacted by either primordial mass segregation or the fraction of binaries in clusters. 

\par \vah{Figure 7 shows that primordial fractality disappears about $5$ Myr earlier in simulations than in the observed cluster sample. This discrepancy is induced by the coevality of stars in the numerical models, which begins immediately after the primordial gas is removed. In contrast, observations indicate that star formation is not strictly instantaneous: although some clusters exhibit very small age spreads ($<0.5$ Myr; \citealt{kud2012} and $<1$ Myr for clusters with mass $<10^3$ M$_\odot$), more massive systems can show spreads of $1-7$ Myr (\citealt{longmore2014rev}). The actual star-forming material propagates along the filaments of progenitor molecular clouds and can persist for up to $\sim 10$ Myr, leading to age spreads of this order within single clusters (\citealt{beccari2010} and \citealt{pang2013}) or adjacent hierarchical structures (\citealt{pang21}). During this phase, fractal stellar substructures can merge and form the final cluster, so fractality disappears on $\sim 10$ Myr timescales in observations, in contrast to the effectively instantaneous star formation assumed in simulations.}

\section{Summary and Conclusion}

\par \vah{In this paper, we analyzed how different initial conditions shape the short-term evolution of young open clusters. For this study, we used eight models with two extreme values of primordial mass segregation (i.e., initially non-segregated and fully segregated), primordial fractality (i.e., initially non-fractalised and fully fractalised), and two different values for the fraction of primordial hard binary systems (i.e., initially with no binaries and with $10 \%$ of hard binary systems). We used the direct $N$-body code \textsc{NBody6++GPU} to dynamically evolve these models up to $100$ Myrs using the standard model for the tidal field of the Galaxy. Our conclusions are the following:}

\begin{enumerate}
\item Our results show that in models with primordial fractality, the inner regions initially undergo contraction and do not exhibit any expansion until roughly $30$ Myr, even when primordial mass segregation is included.

\item In the first Myrs of the cluster life, fractal models exhibit higher average stellar masses in their inner regions compared to models without substructure. This supports the results from (\citealt{mcmillan2007} and \citealt{alis}), which stated that early mass segregation is amplified by the dynamical evolution and merging of primordial substructures during the cluster’s initial development. Furthermore, we find that a fractal model without primordial binaries or mass segregation produces a larger number of massive escapers than any other model.

\item Fractal models generally show better agreement with observational data, especially during the early stages of evolution. The primordial fractal structure dissipates after approximately $5$ Myr, gradually evolving into a more smoothly distributed system. This behavior is consistent with previous studies and matches the contraction period of the inner regions in these models. Our new results suggest that this timescale is largely independent of both the presence of primordial mass segregation or the binary fraction.

\item \vah{The results of comparing the $\Lambda_{MSR}$ values for our models with observed clusters indicate that models without primordial segregation agree more with observations, whereas models with primordial segregation exhibit stronger segregation than typically observed, particularly at early times. Across all models, the presence of primordial binaries has no significant impact on the evolution of mass segregation. However, in models with primordial fractality, they lead to strong but short-lived mass segregation signals during approximately the 10–60 Myr of the simulation.}

\end{enumerate}

\vah{In future work, we will extend this study in several directions to achieve a more comprehensive understanding of the dynamical evolution of young open clusters. Simulations of star clusters with larger numbers of stars and higher stellar densities, along with a broader exploration of primordial fractality, will help to better constrain their impact on cluster evolution. In the present work, our focus regarding binary systems is on comparing models with and without primordial hard binaries, finding that a 10\% fraction of primordial hard binaries is consistent with observational constraints and some previous studies; however, higher fractions of both hard and soft primordial binaries remain to be explored. Extending the simulations to longer timescales, up to several relaxation times, will furthermore allow us to trace the full dynamical evolution of clusters through to their eventual dissolution. Finally, the inclusion of initial cluster rotation may reveal additional dynamical and structural effects not captured in the present models.}

\section*{Acknowledgments}

We thank the anonymous referee for a very detailed and meticulous report, which helped improving the paper significantly.
As a German Academic Exchange Service (DAAD) scholarship holder (Funding Program ID: 57588370), VA acknowledges DAAD for supporting that allowed this study to be conducted. VA, AWHK, and RS also acknowledge NAOC International Cooperation Office for its support in 2023, 2024, and 2025. RS acknowledges Chinese Academy of Sciences President's International Fellowship Initiative for Visiting Scientists (PIFI, grant No. 2026PVA0089), and the National Natural Science Foundation of China (NSFC) under grant No. 12473017. This research was supported in part by the grant NSF PHY-2309135 to the Kavli Institute for Theoretical Physics (KITP).
This material is based upon work supported by Tamkeen under the NYU Abu Dhabi Research Institute grant CASS. FFD, AWHK and RS are grateful for support by the German Science Foundation (DFG), grant No. Sp 345/24-1.
\vah{X.P. acknowledges the financial support of the National Natural Science Foundation of China through grants 12573036 and 12233013, and the China Manned Space Program with grant No. CMS-CSST-2025-A08.}
PB thanks the support from the special program of the Polish Academy of Sciences and the U.S. National Academy of Sciences under the Long-term program to support Ukrainian research teams grant No. PAN.BFB.S.BWZ.329.022.2023. 
BS acknowledges support from the Science Committee of the Ministry of Education and Science, Republic of Kazakhstan (Grant No. AP19677351 and AP13067834) and Nazarbayev University Faculty Development Competitive Research Grant Program (No.  11022021FD2912).
RS acknowledges the support by the National Science Foundation of China (NSFC) under grant No. 12473017. 
 We also acknowledge the Gauss Centre for Supercomputing e.V. for computing time through the John von Neumann Institute for Computing (NIC) on the GCS Supercomputer JUWELS Booster at J\"ulich Supercomputing Centre (JSC) and the use of the parallel computer \texttt{kepler}, originally funded by Volkswagen Foundation. We thank the University Computing Centre Heidelberg (URZ) for support in housing the \texttt{kepler} system.

\section*{Data Availability}
The corresponding author will share the data underlying this article upon reasonable request.


\bibliographystyle{aa}
\bibliography{UCDs}
\end{document}